\documentclass[12pt]{article}

\usepackage{amssymb}
\usepackage{amsmath}
\usepackage{amsfonts}

\newcommand{\R}{\mathbb{R}}

\newcommand{\be}{\begin{equation}}
\newcommand{\bea}{\begin{eqnarray}}
\newcommand{\eea}{\end{eqnarray}}

\newcommand{\kt}{\rangle}
\newcommand{\br}{\langle}

\newcommand{\ed}{\end{document}}

\begin{document}

\title{Pseudo-Hermiticity versus $PT$-Symmetry II: A complete characterization
of non-Hermitian Hamiltonians with a real spectrum}
\author{Ali Mostafazadeh\thanks{E-mail address:
amostafazadeh@ku.edu.tr}\\ \\
Department of Mathematics, Ko\c{c} University,\\
Rumelifeneri Yolu, 80910 Sariyer, Istanbul, Turkey}
\date{ }
\maketitle

\begin{abstract}
We give a necessary and sufficient condition for the reality of the spectrum
of a non-Hermitian Hamiltonian admitting a complete set of biorthonormal eigenvectors.
\end{abstract}

\baselineskip=24pt

Recently, we have explored in \cite{jmp} the basic mathematical structure
underlying the spectral properties of $PT$-symmetric Hamiltonians \cite{pt}.
In particular, we have shown that these properties are associated with a class
 of more general (not necessarily Hermitian) Hamiltonians $H$ satisfying
    \be
    H^\dagger=\eta\,H\,\eta^{-1},
    \label{ps}
    \end{equation}
where $~^\dagger$ denotes the adjoint of the corresponding operator and
$\eta$ is a Hermitian invertible linear operator.
We have termed such a Hamiltonian `$\eta$-{\em pseudo-Hermitian}.' Hermitian
and the $PT$-symmetric Hamiltonians that admit a complete set of biorthonormal
eigenvectors constitute subsets of the set of  pseudo-Hermitian Hamiltonians.
For a $PT$-symmetric Hamiltonian, the exactness of $PT$-symmetry ensures the
reality of the energy spectrum. The purpose of this article is to provide a
complete characterization of the Hamiltonians that have a real spectrum
assuming that they are endowed with a complete set of biorthonormal eigenvectors.

By definition, a $PT$-symmetric Hamiltonian has a symmetry given by an
anti-linear operator, namely $PT$. It is
well-known that if a Hamiltonian satisfies
    \be
    [H,A]=0,
    \label{anti}
    \end{equation}
for an anti-linear operator $A$, then
    \begin{itemize}
    \item[$\star$] either the eigenvalues of $H$ are real or they come
    in complex conjugate pairs.
    \end{itemize}
Furthermore, an eigenvalue of $H$ is real provided that a corresponding
eigenvector is invariant under the action of
$A$, i.e., Eq.~(\ref{anti}) together with
    \be
    H|E\kt=E |E\kt,
    \label{egva}
    \end{equation}
and
    \be
    A|E\kt=|E\kt
    \label{condi-1}
    \end{equation}
imply $E\in\R$. Therefore, a Hamiltonian with an anti-linear symmetry has a
real spectrum if the symmetry is exact.

In Ref.~\cite{jmp}, we have shown that every pseudo-Hermitian Hamiltonian
has the property $\star$. Furthermore, for Hamiltonians with a complete set
of biorthonormal eigenvectors this property is the necessary and sufficient
condition for pseudo-Hermiticity. This, in particular,
means that pseudo-Hermiticity is a necessary condition for having a real
spectrum, but it is not sufficient. In the following we give the necessary
and sufficient condition for the reality of the spectrum of any
Hamiltonian that admits a complete set of biorthonormal eigenvectors. We
shall only consider the case of discrete spectra. The generalization to
continuous spectra does not seem to involve major difficulties.

We first recall the defining properties of a Hamiltonian admitting a
complete set of biorthonormal eigenvectors \cite{bi}. If a Hamiltonian
$H$ has a complete set of biorthonormal eigenvectors
$\{ |\psi_n\kt, |\phi_n\kt \}$, then
    \bea
    &&H |\psi_n\kt=E_n |\psi_n\kt ,~~~~H^\dagger|\phi_n\kt=E_n^*|\phi_n\kt,
    \label{bi-1}\\
    &&\br\phi_m|\psi_n\kt=\delta_{mn},
    \label{bi-2}\\
    &&\sum_n |\psi_n\kt\br\phi_n|=1,
    \label{bi-3}
    \eea
where $n$ is the spectral label, $\delta_{mn}$ denotes the Kronecker
delta function, and $1$ is the identity operator.

    \begin{itemize}
    \item[~]{\bf Theorem:}  Let $H:{\cal H}\to{\cal H}$ be a
    Hamiltonian that acts in a Hilbert space ${\cal H}$,   has a
    discrete spectrum, and admits a complete set of biorthonormal
    eigenvectors $\{ |\psi_n\kt, |\phi_n\kt \}$. Then the spectrum
    of $H$ is real if and only if there is an invertible linear
    operator $O:{\cal H}\to{\cal H}$ such that
    $H$ is $OO^\dagger$-pseudo-Hermitian.
    \item[~]{\bf Proof}: Let $\{|n\kt\}$ be a complete
    orthonormal basis of ${\cal H}$, i.e.,
        \be
        \br m|n\kt=\delta_{mn},~~~~~\sum_{n} |n\kt\br n|=1,
        \label{ortho}
        \end{equation}
    and $O:{\cal H}\to{\cal H}$ and $H_0:{\cal H}\to{\cal H}$ be defined by
        \be
        O:=\sum_{n}|\psi_n\kt\br n|,~~~~~H_0:=\sum_{n}E_n |n\kt\br n|.
        \label{O}
        \end{equation}
    Then, in view of (\ref{bi-1}) -- (\ref{O}), $O$ is invertible with the
    inverse given by
        \be
        O^{-1}= \sum_{n}|n\kt\br \phi_n|,
        \label{e1}
        \end{equation}
    and
        \be
        O^{-1}HO=H_0.
        \label{e2}
        \end{equation}
    Now suppose that the spectrum of $H$ is real. Then, $H_0$ is Hermitian, and
    taking the adjoint of both sides (\ref{e2}), we have
        \be
        O^{-1}HO=O^\dagger H^\dagger {O^{-1}}^\dagger
        \label{e3}
        \end{equation}
    or alternatively
        \be
        H=(OO^\dagger)H^\dagger(OO^\dagger)^{-1}.
        \label{e4}
        \end{equation}
    This equation shows that $H$ is $OO^\dagger $-pseudo-Hermitian. This completes the
    proof of necessity. Next we suppose that $H$ is $OO^\dagger$-pseudo-Hermitian. Then
    (\ref{e4}) and consequently (\ref{e3}) hold. On the other hand, in view of (\ref{bi-2})
    and (\ref{O}), we have
        \[H_0=O^{-1}HO,~~~~~H_0^\dagger=O^\dagger H^\dagger {O^{-1}}^\dagger.\]
    Therefore, (\ref{e3}) implies that $H_0$ is Hermitian, and the
    eigenvalues $E_n$ are all    real.~$\square$
    \end{itemize}
It should be emphasized that the characterization of the non-Hermitian Hamiltonians with a real spectrum given by the preceding theorem applies to the Hamiltonians that admit a complete biorthonormal system of eigenvectors. A generalization of this result to the case of arbitrary Hamiltonians is not known.

\section*{Acknowledgment}
The basic idea of the present work was originated from a comment made by Z.~Ahmed about his
checking the results that I had reported in Ref.~\cite{jmp} for a specific example of a non-Hermitian Hamiltonian with a real spectrum. This project was supported by the Young Researcher Award Program (GEB$\dot{\rm I}$P) of the Turkish Academy of Sciences.

\ed